\documentclass{article}
\usepackage[dvips]{graphicx} 
\usepackage{slashed,booktabs,amsthm}
\usepackage{amsmath}
\usepackage{float,bm,physics,slashed}
\usepackage{amssymb}
\usepackage[caption=false]{subfig}
\usepackage{amsmath}
\usepackage{float}
\usepackage{amssymb}
\usepackage[caption=false]{subfig}
\usepackage[colorlinks=true,linkcolor=blue,filecolor=blue,urlcolor=blue,citecolor=blue]{hyperref}
\usepackage{geometry} 
\usepackage{array,appendix}    
\usepackage{calc}
\usepackage{accents}
\newcommand{\dbtilde}[1]{\accentset{\approx}{#1}}

\usepackage[utf8]{inputenc}
\usepackage{amsmath,mathrsfs}
\usepackage{slashed,csquotes}
\usepackage{xcolor,longtable}
\usepackage{tabularx,colortbl}
\usepackage{matlab-prettifier} 
\usepackage[export]{adjustbox}
\usepackage{graphicx}
\usepackage{csquotes}
\usepackage{array}
\newcommand{\mycomment}[1]{}
\usepackage{tikz}
\usepackage{threeparttable}
\newcolumntype{P}[1]{>{\centering\arraybackslash}p{#1}}
\usepackage{mciteplus}

\begin{document}

\title{Radially Excited Bottom Mesons in Heavy Quark Effective Theory}
\author{Ritu Garg$^1$, Preeti Bhall $^{2*}$, K.K Vishwakarma$^3$ 
\\\small{\it $^{1}$Department of Physics, Manipal University Jaipur, Jaipur }\\\small{\it $^{2,3}$Department of Physics and Material Sciences, Thapar Institute of Engineering and Technology, Patiala}\\\small{E-mail: ritu.garg@jaipur.manipal.edu}, preetibhall@gmail.com}

\date{\today}
\baselineskip =1\baselineskip

\maketitle

\begin{abstract}
 In this paper, Heavy Quark Effective Theory (HQET) is employed to investigate experimentally missing radially excited bottom-meson states. It is an extension of our previous work, in which we had applied HQET for the calculation of charm mesons. By incorporating both theoretical insights and available experimental data on charmed mesons with flavor-symmetry parameters, we computed masses of radially excited bottom-meson states with their strange partners. In addition, decay widths are calculated in the form of hadronic coupling constants $\tilde{\tilde{g}}_{HH}$, $\tilde{\tilde{g}}_{SH}$, $\tilde{\tilde{g}}_{TH}$. By comparing our decay width predictions with available total decay widths, we calculated upper bounds on the corresponding couplings. Regge trajectories are also constructed for our predicted data in planes ($J$, $M^2$ ) and ($n_r$, $M^2$ ), and estimate higher masses ($n = 4$) by fixing Regge slopes and intercepts. Future experimental findings can validate the results obtained in this work.
\end{abstract}

\section{Introduction}
Numerous experimental facilities, including LHCb, BABAR, Belle, BESII/III, CDF, CMS, COMPASS, $D\emptyset$, etc., are constantly involved in producing a lot of hadron data. These experimental facilities provide information on hadron masses, decay widths, branching ratios, spin, parity, polarization amplitudes, etc. Recently, LHCb detected new excited charmed strange mesons $D_{s0}(2590)^{+}$ in the decay process $B^{0} \longrightarrow D^{-}D^{+}K^{+}\pi^{-}$ with large statistical error \cite{aaij2021}. The measured mass and decay width are:
\begin{center}
    $M(D_{s0}(2590))= 2591\pm6(stat)\pm7(syst) MeV$\\
    $\Gamma(D_{s0}(2590)) = 89\pm16\pm12$ $MeV$
\end{center}
    Furthermore, LHCb identified this state as radial excitation of the ground state, that is, $n = 2$, $L = 0$. Numerous theoretical models have been studied to uncover the nature of the charmed strange meson $D_{s0}(2590)^{+}$. The semi-relativistic potential model \cite{2}, PCAC low energy theory \cite{1}, and coupled channel framework \cite{4} were employed to analyze the state $D_{s0}(2590)^{+}$. These studies suggested that to incorporate $D_{s0}(2590)^{+}$ into charm spectra, further observations are required in future investigations. In the charm- meson sector, finding of some more excited states like $D_{0}(2550)$, $D_{1}^{*}(2600)$, $D_{2}(2740)$, $D_{3}^{*}(2760)$, $D_{J}(3000)^{0}$, $D_{J}^{*}(3000)$ and strange states $D_{s1}(2860)$, $D_{sJ}(3040)$, 
$D_{s0}(2590)$ \cite{5,6,7,8,9,10}  not only widened the spectra but also facilitated the exploration of their properties through decay analyses. Several theoretical models, including the $^3P_{0}$ model \cite{12}, HQET \cite{13}, QCD sum rule \cite{14}, and relativized quark model \cite{15,16}, analyzed all of the above states, calculated their masses and presented their $J^{P}$ values.

However, experimental progress toward establishing the bottom sector is still limited. Only ground states $B^{0,\pm}(5279)$, $B^{*}(5324)$, $B_{s}(5366)$, $B_{s}^{*}(5415)$ and some low-lying states $B_{1}(5721)$, $B_{J}^{*}(5732)$, $B_{2}^{*}(5747)$, $B_{s1}(5830)$, $B_{s2}^{*}(5840)$, $B_{sJ}^{*}(5850)$, $B_{J}(5840)$, $B_{J}(5970)$ are experimentally observed and enumerated in PDG \cite{11}. However, apart from these states, the whole bottom-meson spectrum is unknown. Both experimental efforts and theoretical models are actively engaged in finding new states that could help bridge this gap and complete the bottom-meson spectrum. In this process, recently, LHCb collaborations identified two new states $B_{sJ}(6063)$ and $B_{sJ}(6114)$ in the $B^{+}K^{-}$ mass spectrum \cite{30}. The measured masses and decay widths are given below:\\
\begin{center}
   $M(B_{sJ}(6063))= 6063.5\pm1.2(stat)\pm0.8(syst) MeV$\\
$\Gamma(B_{sJ}(6063)) = 26\pm4\pm4$ $MeV$
\end{center}
\begin{center}
$M(B_{sJ}(6114))= 6114\pm3(stat)\pm5(syst) MeV$\\
$\Gamma(B_{sJ}(6114)) = 66\pm18\pm21$ $MeV$ 
\end{center}

These newly announced bottom meson states by LHCb have opened the window to investigate and enhance our understanding of higher excited bottom states.

In theory, various theoretical studies have performed different analyses for higher excited bottom non-strange and bottom strange meson states \cite{13a,14a,15a,16a,17a,18a,19a,20a,21a,22a,23a,24a,25a,26a,27a,28a,29a,30a,31a,32a,33a}. By employing theoretical models, the states $B^{0,\pm}(5279)$, $B^{*}(5324)$, $B_{s}(5366)$, $B_{s}^{*}(5415)$, identified as states $1S$, consistent with experimental measurements. Furthermore, the states $B_{1}(5721)$ and $B_{2}^{*}(5747)$ are also experimentally confirmed and categorized as $1P$ states with quantum numbers $1^+$ and $2^+$, respectively. However, theoretically, the state $B_{1}(5721)$ remains an ambiguous candidate. This is since some theoretical work like heavy meson effective theory supports it as a $1P(1^+)$ state \cite{25a,32a}, while other work using the relativistic quark model and non-relativistic quark model explains it as a combination of $^3P_1$ and $^1P_1$ states.\cite{14a,16a,33a}. The $J^P$ of the state $B_J(5840)$ is still not confirmed, as different models suggest a different $J^P$ for it. The authors in \cite{16a,17a} analyzed the $B_J(5840)$ state using the quark model and proposed its assignment as the $2^1S_0$ state. In contrast, G. L. Yu and Z. G. Wang, based on an analysis within the $^3P_0$ decay model framework, favor its classification as the $2^3S_1$ state \cite{20a}. However, the effective theory of the heavy quark explained resonances $B_J(5840)$ as $1^3D_1$ state \cite{28a}. The state $B_J(5960)^{0,+}$ was interpreted differently in various theoretical models, with proposed assignments including states $2^3S_1$, $1^3D_3$, or $1^3D_1$ \cite{14a,15a,16a,22a,23a,24a,34a}. But its $J^P$ is still a question mark in the PDG, which only mentions its mass and decay width. We discussed a short literature review on these non-strange bottom states. The assignments of these states ($B_{1}(5721)$, $B_{2}^{*}(5747)$, $B_J(5970)$) are also suggested in our previous work \cite{28a}. In the strange bottom sector, only a few states have been observed to date. Among them, the states $B_{s1}(5830)$ and $B_{s2}(5840)$ have been clearly identified by the CDF, D0, and LHCb collaborations, and are classified as the $1P$ states with quantum numbers $1^+$ and $2^+$, respectively \cite{11}. However, there is ambiguity regarding the recently observed strange bottom meson states $B_{sJ}(6063)$ and $B_{sJ}(6114)$. The states $B_{sJ}(6063)$ and $B_{sJ}(6114)$ are identified within the non-relativistic quark potential model as the $1^3D_1$ and $1^3D_3$ states, respectively. However, the authors in Ref. \cite{34a} assign these states to the $1^3D_1$ and $2^3S_1$ states, respectively. Theoretical analysis of these newly observed states remains limited in the current literature, highlighting the need for further investigation.

\begin{table*}
\caption{\label{differentcolloborationsobservationsforbotomm} Masses and Decay widths of bottom and bottom strange mesons observed by various collaborations.}
 \centering
       \begin{tabular} {|c|c|c|c|c|}
           \hline
          State & $J^{P}$ & Mass ($MeV$) & Width ($MeV$) & Experiment  \\
          \hline
          $B(5279)^{0}$ & $0^{-}$ & $5279.61\pm{0.16}$ & - & CLEO \cite{behrends1983} \\
          \hline
           $B(5279)^{\pm}$ & $0^{-}$ & $5279.61\pm{0.15}$ & - & CLEO \cite{behrends1983} \\
           \hline
            $B(5324)^{*}$ & $1^{-}$ & $5324.83\pm{0.32}$ & - & CUSB \cite{han1985} \\
            \hline
            $B_{s}(5366)$ & $0^{-}$ & $5366.79\pm{0.23}$ & - & CUSB-II \cite{lee1990} \\
            \hline
            $B_{s}^{*}(5415)$ & $1^{-}$ & $5415.40\pm{0.15}$ & - & CUSB-II \cite{lee1990}  \\
            \hline
             $B_{J}(5721)$ & $1^{+}$ & $5727.7\pm{0.7}$ & $30.1\pm{1.5}$ & LHCb \cite{aaij2015b} \\
             
                 &   & $5720.6\pm{0.24}$ &   & $DO$ \cite{abazov2007}  \\
                 
                 &  & $5725.3\pm{1.6}$ &  & CDF \cite{aaltonen2014} \\
                 \hline
                  $B_{2}^{*}(5747)$ & $2^{+}$ & $5739.44\pm{0.37}$ & $24.5\pm 1.0$ & LHCb \cite{aaij2015b} \\
                  
                    &  & $5746.8\pm{0.24}$ &  & $DO$ \cite{abazov2007} \\
                    
                     &  & $5740.2\pm{1.7}$ & $22.7\pm{32}$ & CDF \cite{aaltonen2009} \\
                     \hline
                       $B_{s1}(5830)$ & $1^{+}$ & $5828.40\pm{0.04}$ & - & LHCb \cite{aaij2013} \\
                      
                        &  & $5828.3\pm{0.1}$ & $0.5\pm{0.3}$ & CDF \cite{aaltonen2014}  \\
                        
                         &  & $5829.4\pm{0.7}$ &  & CDF \cite{aaltonen2008} \\
                         \hline
                         $B_{s2}(5840)$ & $2^{+}$ & $5839.60\pm{1.1}$ & - & $DO$ \cite{abazov2008}  \\
                         
                           &  & $5839.
                          70\pm{0.7}$ & - & CDF \cite{aaltonen2008}  \\
                           
                            &  & $5839.
                          70\pm{0.1}$ & $1.40\pm{0.4}$ & CDF \cite{aaltonen2014} \\
                          
 &  & $5839.
                          99\pm{0.05}$ & $1.56\pm{0.13}$ & LHCb \cite{aaij2013}   \\
                          \hline 
                       $B_{J}(5840)$ & - & $5862.9\pm{5.0}$ & $127.4\pm{16.7}$ & LHCb\cite{aaij2015b} \\
                       \hline
                       $B_{J}(5960)$ & - & $5978.9\pm{5.0}$ & - & CDF \cite{aaltonen2009}  \\
                       
                         & - & $5969.2\pm{2.9}$ & $82.3\pm7.7$ & LHCb \cite{aaij2015b}\\
                         \hline
                      $B_{J}(5970)^{0}$ & - & $5961\pm{5.0}$ & $60\pm{40}$ & CDF \cite{aaltonen2014} \\
                      \hline
                      $B_{sJ}(6064)$ &- & $6063.5 \pm{1.4}$ & $26 \pm {6}$& LHCb \cite{aaij2021}\\
                      \hline
        $B_{sJ}(6114)$ & & $6114 \pm{6}$ & $66 \pm {6}$& LHCb \cite{aaij2021}\\ 
        \hline

\end{tabular}

\end{table*}

A review of the existing literature reveals that experimental observations of higher orbitally and radially excited bottom-meson states, including their strange counterparts, are still lacking to date. The present theoretical investigation is motivated by the lack of experimental data on several bottom meson states, particularly their strange partners. In this work, we investigate the properties of these experimentally missing bottom mesons, including their masses, decay widths, and upper boundaries on hadronic coupling constants, within the framework of heavy-quark effective theory (HQET). This paper is organized as follows: Section 2 outlines the HQET framework employed to analyze strong decay processes. In Section 3, we present and discuss the results of our calculations. Finally, Section 4 summarizes the conclusions of this study.

\section{Framework}
Heavy-light hadrons can be effectively studied within the framework of Heavy Quark Effective Theory (HQET). HQET serves powerful framework for describing properties of heavy-light mesons like masses, decay widths, branching ratios, fractions, spin, parity, etc \cite{grinstein1990, neubert1996}. This theory is formulated based on two approximate symmetries: heavy quark symmetry and chiral symmetry. Heavy quark symmetry holds in the limit where the heavy quark mass approaches infinity, $m_Q\to\infty$. In the limit $m_Q\rightarrow \infty$, the spin of the light degrees of freedom decouples from the spin of the heavy quark. As a result, the total angular momentum of the light components remains conserved \cite{georgi1990, eichten1989}. The total angular momentum of the light degrees of freedom is given by $s_{l} = s_{q} + l$, $s_{q}$ is the spin of light quark (1/2), and $l$ is the total orbital momentum of light quarks. In the heavy quark limit, mesons are arranged in doublets based on the total angular momentum $s_{l}$ of light quarks. For $l = 0$, $s_{l} = 1/2$ coupled this with spin of heavy quark $s_{Q}$ = 1/2 and resulted with doublet $(0^{-},1^{-})$. These doublets are denoted by $(P, P^{*})$. For $l = 1$, two doublets are formed based on the total angular momentum of the light degrees of freedom. These are denoted by $(P^{*}_{0}, P_{1}^{'})$ and $(P_{1}, P_{2}^{*})$ corresponding to $J_{s_{l}}^{P} = (0^{+}, 1^{+})$ and $J_{s_{l}}^{P} = (1^{+}, 2^{+})$ respectively. For $l = 2$, two doublets are formed: $(P^{*}_{1}, P_{2})$ and $(P_{2}^{'},P_{3}^{*})$ corresponding to $J_{s_{l}}^{P} = (1^{-}, 2^{-})$ and $J_{s_{l}}^{P} = (2^{-}, 3^{-})$ respectively. Similarly for $l = 3$, doublets $(P^{*}_{2}, P_{3})$ and $(P_{3}^{'},P_{4}^{*})$ are formed associated with $J_{s_{l}}^{P} = (1^{-}, 2^{-})$ and $J_{s_{l}}^{P} = (2^{-}, 3^{-})$ respectively. These doublets are described using super effective fields $H_{a}, S_{a}, T_{a}, X^{\mu}_{a}, Y^{\mu\nu}_{a}$ and expression for fields are given below:
\begin{gather}
\label{eq:lagrangian}
 H_{a}=\frac{1+\slashed
v}{2}\{P^{*}_{a\mu}\gamma^{\mu}-P_{a}\gamma_{5}\}\\
S_{a} =\frac{1+\slashed v}{2}[{P^{'\mu}_{1a}\gamma_{\mu}\gamma_{5}}-{P_{0a}^{*}}]\\
T^{\mu}_{a}=\frac{1+\slashed v}{2}
\{P^{*\mu\nu}_{2a}\gamma_{\nu}-P_{1a\nu}\sqrt{\frac{3}{2}}\gamma_{5}
[g^{\mu\nu}-\frac{\gamma^{\nu}(\gamma^{\mu}-\upsilon^{\mu})}{3}]\}
\end{gather}
\begin{gather}
X^{\mu}_{a}=\frac{1+\slashed
v}{2}\{P^{\mu\nu}_{2a}\gamma_{5}\gamma_{\nu}-P^{*}_{1a\nu}\sqrt{\frac{3}{2}}[g^{\mu\nu}-\frac{\gamma_{\nu}(\gamma^{\mu}+v^{\mu})}{3}]\}
\end{gather}

\begin{multline}
 Y^{\mu\nu}_{a}=\frac{1+\slashed
v}{2}\{P^{*\mu\nu\sigma}_{3a}\gamma_{\sigma}-P^{'\alpha\beta}_{2a}\sqrt{\frac{5}{3}}\gamma_{5}[g^{\mu}_{\alpha}g^{\nu}_{\beta}
-\frac{g^{\nu}_{\beta}\gamma_{\alpha}(\gamma^{\mu}-v^{\mu})}{5}-\frac{g^{\mu}_{\alpha}\gamma_{\beta}(\gamma^{\nu}-v^{\nu})}{5}]\}   
\end{multline}
The field $H_a$ represents doublets of S-wave with $J^P = (0^-,1^-)$. The fields $S_a$ and $T_a$ describes doublets of P-wave for $J^P = (0^+, 1^+)$ and $(1^+, 2^+)$ respectively. D-wave doublets with $J^P = (1^-, 2^-)$ and $(2^-, 3^-)$ described by $X^{\mu}_{a}$ and $Y^{\mu\nu}_{a}$ respectively. $a$  is light quark ($u, d, s$) flavor index. The symbol $v$  denotes the four-velocity of the heavy quark, which remains conserved in strong interaction processes. The approximate chiral symmetry $SU(3)_L\times SU(3)_R$ is involved with fields of pseudoscalar mesons $\pi$, K, and $\eta$ which are lightest strongly interacting  bosons \cite{cho1992, kilian1992}. They are treated as approximate Goldstone bosons of this chiral symmetry and can be introduced by the matrix field $U(x) = Exp\left[\dot{\iota}\sqrt{2}\phi(x)/f\right]$, where $\phi(x)$ is given by
\begin{align}
\phi(x) =
\begin{pmatrix}
\frac{1}{\sqrt{2}}\pi^0+\frac{1}{\sqrt{6}}\eta & \pi^+ & K^+\\
\pi^-& -\frac{1}{\sqrt{2}}\pi^0+\frac{1}{\sqrt{6}}\eta & K^0\\
K^- & \bar{K}^0 & -\sqrt{\frac{2}{3}}\eta
\end{pmatrix}
\end{align}
The fields of heavy meson doublets (1-7) interact with pseudoscalar goldstone bosons via covariant derivative $D_{\mu ab}= -\delta_{ab}\partial_{\mu}+\mathcal{V}_{\mu ab} =  -\delta_{ab}\partial_{\mu}+\frac{1}{2}(\xi^{+}\partial_{\mu}\xi+\xi\partial_{\mu}\xi^{+})_{ab}$ and axial vector field $A_{\mu ab}=\frac{i}{2}(\xi\partial_{\mu}\xi^{\dag}-\xi^{\dag}\partial_{\mu}\xi)_{ab}$. By incorporating all meson doublet fields along with the Goldstone boson fields, the effective Lagrangian is expressed as \cite{scherer2002, davoudiasl1996, wise1992}:
\begin{multline}
    \mathcal{L} = iTr[\bar{H}_{b}v^{\mu}D_{\mu ba}H_{a}] +  \frac{f_\pi^{2}}{8}Tr[\partial^{\mu}\Sigma\partial_{\mu}\Sigma^{+}] + Tr[\bar{S_{b}}(iv^{\mu}D_{\mu ba} - \delta_{ba}\Delta_{S})S_{a}]+Tr[\bar{T_{b}^{\alpha}}(iv^{\mu}D_{\mu ba}- \delta_{ba}\Delta_{T})\\T_{a \alpha}- Tr[\bar{X_{b}^{\alpha}}(iv^{\mu}D_{\mu ba}- \delta_{ba}\Delta_{X})X_{a \alpha}+ Tr[\bar{Y_{b}^{\alpha\beta}}(iv^{\mu}D_{\mu ba}- \delta_{ba}\Delta_{Y})Y_{a\alpha\beta}] 
\end{multline}
The mass parameter $\Delta_{F}$ in equation (9) presents the mass difference between higher mass doublets ($F$) and the lowest lying doublet ($H$) in terms of spin average masses of these doublets with the same principal quantum number ($n$). The expressions for mass parameters are given by:
\begin{subequations}
\label{averagemassequation}
\begin{align}
     \Delta_{F}=\overline{M_{F}}&- \overline{M_{H}},~~ F= S,T,X,Y\\
\text{where, } \quad
 \overline{M_{H}}&=(3m^{Q}_{P_1^*}+m^{Q}_{P_{0}})/4\\
\overline{M_{S}}&=(3m^{Q}_{P_1^{'}}+m^{Q}_{P_0^*})/4\\
\overline{M_{T}}&=(5m^{Q}_{P_2^*}+3m^{Q}_{P_1})/8\\
\overline{M_{X}}&=(5m^{Q}_{P_2}+3m^{Q}_{P_1^{*}})/8\\
 \overline{M_{Y}}&=(5m^{Q}_{P_3^*}+3m^{Q}_{P_2^{'}})/8
 \end{align}
 \end{subequations}
           The symmetry-breaking terms account for the $1/m_{Q}$ corrections to the heavy quark limit. The corrections take the form of: 
           \begin{multline}
          \mathcal{L}_{1/m_{Q}} = \frac{1}{2m_{Q}}[\lambda_{H} Tr(\overline H_{a}\sigma^{\mu\nu}{H_{a}}\sigma_{\mu\nu}) + \lambda_{S}Tr(\overline S_{a}\sigma^{\mu\nu} S_{a}\sigma_{\mu\nu})+\lambda_{T}Tr(\overline T_{a}^{\alpha}\sigma^{\mu\nu}{T_{a}^{\alpha}}\sigma_{\mu\nu})]\\+\lambda_{X}Tr(\overline X_{a}^{\alpha}\sigma^{\mu\nu}{X_{a}^{\alpha}}\sigma_{\mu\nu}) + \lambda_{Y}Tr(\overline Y_{a}^{\alpha\beta}\sigma^{\mu\nu}{Y_{a}^{\alpha\beta}}\sigma_{\mu\nu})]
          \end{multline} 
          The parameters $\lambda_{H}$, $\lambda_{S}$, $\lambda_{T}$, $\lambda_{X}$, and $\lambda_{Y}$ are corresponding to hyperfine splittings and are defined as specified in Eq. \eqref{lambdaparameters}. The mass terms in the Lagrangian represent first-order contributions in $1/m_{Q}$, while higher-order factors may also be present. We are restricting our analysis to the first-order corrections in $1/m_{Q}$.
\begin{subequations}
 \label{lambdaparameters}
    \begin{gather}
     \lambda_{H} = \frac{1}{8}(M^{2}_{P^{*}} - M^{2}_{P}) \\
     \lambda_{S} = \frac{1}{8}({M^{2}_{P_1^{'}}} - {M^{2}_{P_0^*}})\\
     \lambda_{T}=\frac{3}{8}({M^{2}_{P_2^*}}-{M^{2}_{P_1}})\\
     \lambda_{X}=\frac{3}{8}({M^{2}_{P_2}}-{M^{2}_{P_1^*}})\\
      \lambda_{Y}=\frac{3}{8}({M^{2}_{P_3}}-{M^{2}_{P_2^{'*}}})
      \end{gather}
\end{subequations}
  Here, we are motivated by the fact that at an energy scale of approximately 1 GeV, the study of Heavy Quark Effective Theory (HQET) reveals an emergent flavor symmetry for the bottom ($b$) and charm ($c$) quarks. As a result, the following relationships reflect the elegance of flavor symmetry:

       \begin{subequations}
  \label{parametersymmetry}
     \begin{align}
          \Delta_{F}^{(c)} =\Delta_{F}^{(b)}\\
           \lambda_{F}^{(c)} = \lambda_{F}^{(b)}
   \end{align}
 \end{subequations}

The decays $F\rightarrow  H + G$ (F = H, S, T, X, Y
, G representing a light pseudoscalar meson) can be described using effective Lagrangians, which are explained in terms of the fields defined in (9-14). These Lagrangians are applicable at leading order in both the heavy quark mass expansion and the light meson momentum expansion \cite{casalbuoni1992}:
\begin{gather}
\label{eq:lagrangian1}
L_{HH}=g_{HH}Tr\{\overline{H}_{a}
H_{b}\gamma_{\mu}\gamma_{5}A^{\mu}_{ba}\}\\
L_{TH}=\frac{g_{TH}}{\Lambda}Tr\{\overline{H}_{a}T^{\mu}_{b}(iD_{\mu}\slashed
A + i\slashed D A_{\mu})_{ba}\gamma_{5}\}+h.c.\\
L_{XH}=\frac{g_{XH}}{\Lambda}Tr\{\overline{H}_{a}X^{\mu}_{b}(iD_{\mu}\slashed
A + i\slashed D A_{\mu})_{ba}\gamma_{5}\}+h.c.
\end{gather}
\begin{multline}
L_{YH}=\frac{1}{\Lambda^{2}}Tr\{\overline{H}_{a}Y^{\mu\nu}_{b}[k^{Y}_{1}\{D_{\mu}
,D_{\nu}\}A_{\lambda}+k^{Y}_{2}(D_{\mu}D_{\lambda}A_{\nu}
+D_{\nu}D_{\lambda}A_{\mu})]_{ba}\gamma^{\lambda}\gamma_{5}\}+h.c.
\end{multline}
In these equations $D_{\mu} =
\partial_{\mu}+V_{\mu}$,  $\{D_{\mu},D_{\nu}\}
= D_{\mu}D_{\nu}+D_{\nu}D_{\mu}$ and $\{D_{\mu} ,D_{\nu}D_{\rho}\} =
D_{\mu}D_{\nu}D_{\rho}+D_{\mu}D_{\rho}D_{\nu}+D_{\nu}D_{\mu}D_{\rho}+D_{\nu}D_{\rho}D_{\mu}+D_{\rho}D_{\mu}
D_{\nu}+D_{\rho}D_{\nu}D_{\mu}$. $\Lambda$ is the chiral symmetry breaking scale taken as 1 GeV. $g_{HH}$, $g_{SH}$,  $g_{TH}$, $g_{XH}$, $g_{YH}
= k^{Y}_{1}+k^{Y}_{2}$ are the
strong coupling constants involved. The interactions between the ground state positive and negative parity bottom mesons and the higher excited bottom states, as well as the emission of light pseudo-scalar mesons $(\pi, \eta, K)$, are explained by these equations. The strong decays of two bodies are calculated using the Lagrangians $L_{HH}, L_{SH}, L_{TH}, L_{XH}, L_{YH}$, and expressions for decays are given by \cite{wang2014, colangelo2012}:
\\$(0^{-},1^{-}) \rightarrow (0^{-},1^{-}) + G$
\begin{eqnarray}
\label{eq:3.13} \Gamma(1^{-} \rightarrow 1^{-})=
C_{G}\frac{g_{HH}^{2}M_{2}p_{G}^{3}}{3\pi
f_{\pi}^{2}M_{1}}\\\nonumber \Gamma(1^{-} \rightarrow 0^{-})=
C_{G}\frac{g_{HH}^{2}M_{2}p_{G}^{3}}{6\pi
f_{\pi}^{2}M_{1}}\\\nonumber \Gamma(0^{-} \rightarrow 1^{-})=
C_{G}\frac{{g}_{HH}^{2}M_{2}p_{G}^{3}}{2\pi
f_{\pi}^{2}M_{1}}
\end{eqnarray}

 $(0^{+},1^{+}) \rightarrow (0^{-},1^{-}) + G$
\begin{eqnarray}
\label{eq:3.14} \Gamma(1^{+} \rightarrow 1^{-})=
C_{G}\frac{g_{SH}^{2}M_{2}(p^{2}_{G}+m^{2}_{G})p_{G}}{2\pi
f_{\pi}^{2}M_{1}}\\\nonumber \Gamma(0^{+} \rightarrow 0^{-})=
C_{G}\frac{g_{SH}^{2}M_{2}(p^{2}_{G}+m^{2}_{G})p_{G}}{2\pi
f_{\pi}^{2}M_{1}}
\end{eqnarray}

 $(1^{+},2^{+}) \rightarrow (0^{-},1^{-}) + G$
\begin{eqnarray}
\label{eq:3.15} \Gamma(2^{+} \rightarrow 1^{-})=
C_{G}\frac{2g_{TH}^{2}M_{2}p_{G}^{5}}{5\pi
f_{\pi}^{2}\Lambda^{2}M_{1}}\\\nonumber \Gamma(2^{+} \rightarrow
0^{-})= C_{G}\frac{4g_{TH}^{2}M_{2}p_{G}^{5}}{15\pi
f_{\pi}^{2}\Lambda^{2}M_{1}}\\\nonumber \Gamma(1^{+} \rightarrow
1^{-})= C_{G}\frac{2g_{TH}^{2}M_{2}p_{G}^{5}}{3\pi
f_{\pi}^{2}\Lambda^{2}M_{1}}
\end{eqnarray}
$(1^{-},2^{-}) \rightarrow (0^{-},1^{-}) + G$
\begin{eqnarray}
\label{eq:3.16} \Gamma(1^{-} \rightarrow 0^{-})=
C_{G}\frac{4g_{XH}^{2}}{9\pi f_{\pi}^{2}\Lambda^{2}}
\frac{M_{2}}{M_{1}}[p_{G}^{3}(m_{G}^{2}+p_{G}^{2})]\\\nonumber
\Gamma(1^{-} \rightarrow 1^{-})= C_{G}\frac{2g_{XH}^{2}}{9\pi
f_{\pi}^{2}\Lambda^{2}}
\frac{M_{2}}{M_{1}}[p_{G}^{3}(m_{G}^{2}+p_{G}^{2})]\\\nonumber
\Gamma(2^{-} \rightarrow 1^{-})= C_{G}\frac{2g_{XH}^{2}}{3\pi
f_{\pi}^{2}\Lambda^{2}}
\frac{M_{2}}{M_{1}}[p_{G}^{3}(m_{G}^{2}+p_{G}^{2})]
\end{eqnarray}
In the above expressions, $M_{1}$, $M_{2}$ represents initial and final meson masses, $\Lambda$ = 1 $GeV$ is chiral symmetry breaking scale. $p_{G}$, $m_{G}$ is the final momentum and mass of the light pseudoscalar meson. The phenomenology study of heavy-light mesons is strongly influenced by the coupling constant. These dimensionless coupling constants $g_{HH}$, $g_{SH}$, $g_{TH}$, etc, provide the strength of transition between different heavy-light mesons fields such as $H-H$ field (negative-negative parity), $S-H$ field (positive-negative parity), $T-H$ field (positive-negative parity). For transition involving radial excitations from $n = 3$ to $n = 1$ coupling constants are given by $\tilde{\tilde{g}}_{HH}$, $\tilde{\tilde{g}}_{SH}$, $\tilde{\tilde{g}}_{TH}$. The coefficient $C_{G}$ for different pseudoscalar particles are:
$C_{\pi^{\pm}}$, $C_{K^{\pm}}$, $C_{K^{0}}$, $C_{\overline{K}^{0}}=1$, $C_{\pi^{0}}=\frac{1}{2}$ and $C_{\eta}=\frac{2}{3}(c\bar{u}, c\bar{d})$ or $\frac{1}{6}(c\bar{s})$. Higher order corrections of $\frac{1}{m_{Q}}$ for spin and flavor violations are not included in bringing new couplings.
\section{Results and Discussions}
New findings in the bottom sector like $B_{J}(5840)^{0,\pm}$, $B_{J}(5960)$, $B_{J}(5970)$ and strange states $B_{s2}^{*}(5840)$, $B_{sJ}(6064)$, $B_{sJ}(6114)$ have enriched bottom spectrum \cite{PDG2022}. Despite the discovery of these new states, the experimental exploration of higher bottom and bottom-strange meson spectra remains limited. In this section, we have investigated the higher bottom and bottom-strange meson sector and calculated the masses, decay widths, and corresponding coupling constants for the $n = 3$ radial excitation within the framework of Heavy Quark Effective Theory (HQET). We first calculated the masses and subsequently predicted the decay widths using these results, in order to assess the reliability of the HQET model. We organized our analysis into two parts: in the first, we predicted the masses of bottom mesons for $n = 3$, along with their strange partners; in the second, we calculated the strong decay widths in terms of coupling constants, using the predicted masses as input.

\subsection{Masses}
Mass is a fundamental parameter in describing the spectroscopy of heavy-light mesons. The input values used for calculating the masses of bottom and bottom-strange mesons are listed in Table \ref{tab:inputforbottom}

\begin{table}[ht]
\centering
\caption{Input values $3S$ taken from Ref. \cite{ebert2010} and remaining are from \cite{Garg2022charm}. All values are in units of $MeV$.}
\begin{tabular}{|c|c|c|c|c|c|}
\hline
State & $J^{P}$ & $ c\overline{q}$ & $ c\overline{s}$ & $ b\overline{q}$ & $ b\overline{s}$ \\
\hline
 $3^{1}S_{0}$ & $0^{-}$ & 3062 \cite{ebert2010} & 3219 \cite{ebert2010} & - & -  \\
\hline
$3^{3}S_{1}$ & $1^{-}$ & 3096 \cite{ebert2010} & 3242 \cite{ebert2010} & - & - \\
\hline
$3^{3}P_{0}$ & $0^{+}$ & 3243.17 & 3496.46 & - & -  \\
\hline
$3^{1}P_{1}$ & $1^{+}$ & 3356.13 & 3367.18 & - & -  \\
\hline
$3^{3}P_{1}$ & $1^{+}$ & 3281.27  & 3508.63 & - & -  \\
\hline
$3^{3}P_{2}$ & $2^{+}$ &3337.81  & 3563.20 & - & -  \\
\hline
$2^{1}S_{0}$ & $0^{-}$ & 2581 &2688& 5890 &  5976   \\
\hline
$2^{3}S_{1}$ & $1^{-}$ & 2632 &  2731& 5906 & 5992  \\
\hline
\end{tabular}
\label{tab:inputforbottom}
\end{table}

To compute the masses of $n = 3 $ bottom meson states, we first calculated values of average masses $ \overline{M_{\Tilde{\Tilde{H}}}}$, $ \overline{M_{\Tilde{\Tilde{S}}}}$, $ \overline{M_{\Tilde{\Tilde{T}}}}$  ($\approx$ labels for $n =3$) introduced in Eqs. \ref{averagemassequation} for charm meson states from Table \ref{tab:inputforbottom}, then HQS (heavy symmetry parameters) $\Delta_{\Tilde{\Tilde{H}}}$, $\Delta_{\Tilde{\Tilde{S}}}$, $\Delta_{\Tilde{\Tilde{T}}}$, $\lambda_{\Tilde{\Tilde{H}}}$, $\lambda_{\Tilde{\Tilde{S}}}$, and  $\lambda_{\Tilde{\Tilde{T}}}$ described in Eqs.\ref{lambdaparameters} are calculated for same charm meson states. Since $\Delta_{F}$, $\lambda_{F}$ are flavor independent in HQET, which implies  $\Delta_{F}^{(b)} =\Delta_{F}^{(c)}$, $\lambda_{F}^{(b)} = \lambda_{F}^{(c)}$. With the calculated symmetry parameters  $\Delta_{F}$, $\lambda_{F}$ for charm meson states and then applying heavy quark symmetry, we predicted the masses for $ n=3$ bottom mesons listed in Table \ref{massresultsforbottommesons}. For the details, we elaborate on the calculation part of the mass $B(3^1 S_{0})$. From Table \ref{tab:inputforbottom}, using charm states, we calculated $ \overline{M^{c}_{\Tilde{\Tilde{H}}}} = 3087.50$ $MeV$, $ \overline{M^{c}_{\Tilde{H}}} = 2619.25$ $MeV$. Then, using these two values, we have $\Delta^{c}_{\Tilde{\Tilde{H}}} = \overline{M^{c}_{\Tilde{\Tilde{H}}}} - \overline{M^{c}_{\Tilde{H}}} =  468.25$ $MeV$. Using Eq. \ref{lambdaparameters}, we get $\lambda^{c}_{\Tilde{\Tilde{H}}} = 26171.50 $ $MeV^{2}$. The symmetry of these parameters given by Eq. \ref{parametersymmetry} and implies that $\Delta^{b}_{\Tilde{\Tilde{H}}} = 468.25$ $MeV$, $\lambda^{b}_{\Tilde{\Tilde{H}}} = 26171.50 $ $MeV^{2}$. Using the values of $\Delta^{b}_{\Tilde{\Tilde{H}}} = 468.25$ $MeV$ and $\lambda^{b}_{\Tilde{\Tilde{H}}} = 26171.50 $, we obtained the mass of $B(3^1 S_{0}) = 6326.22$ $MeV$. Similarly, by following the same procedure, we obtained other masses for $n = 3$ bottom mesons listed in Table \ref{massresultsforbottommesons}. As we know, experimental observations for higher excited states concerning $B$-mesons are fewer compared to $D$-mesons. It motivates us to explore more bottom spectra and check the validity of our theoretical model (HQET).

On comparison, our calculated masses for $n = 3$ bottom meson states are in good agreement with other theoretical estimates for both strange and non-strange states. Our results are nicely with masses obtained by the relativistic quark model \cite{ebert2010} with a deviation of $\pm 1\%$. On comparing with Ref. \cite{godfrey2016b}, our results are deviated by $\pm 2\%$.

\begin{table}[ht!]
\centering

\caption{Predicted masses for radially excited bottom mesons.}
 \begin{tabular}{| c | c | c | c | c | c | c |}
      \hline
      \multicolumn{1}{|c|}{} & \multicolumn{6}{c|}{Masses of $n = 3$ bottom Mesons ($MeV$)}\\
       \cline{2-7}
     
       \multicolumn{1}{|c|}{$J^{P}(n^{2S+1}L_{J})$}&\multicolumn{3}{c|}{Non-Strange}&\multicolumn{3}{c|}{Strange}\\
       \cline{2-7}
       & \multicolumn{1}{c|}{Present Results}&\multicolumn{1}{c|}{\cite{ebert2010}}&\multicolumn{1}{c|}{\cite{godfrey2016b}}&\multicolumn{1}{c|}{Present Results}&\multicolumn{1}{c|}{\cite{ebert2010}}&\multicolumn{1}{c|}{\cite{godfrey2016b}}\\
       \hline
        $0^{-}(3^1S_0)$ &  6326 &  6379 & 6335& 6463 & 6467 & 6301\\
        \hline
        $1^{-}(3^3S_1)$ &  6342 &  6387 &6355   & 6474 & 6475 & 6319 \\
        \hline
        $0^{+}(3^3P_0)$ &  6568& 6629  & 6576& 6789 & 6731 & 6504\\
        \hline
        $1^{+}(3^1P_1)$ &  6624 &  6685 &  6585& 6826 & 6768 & 6519\\
        \hline
        $1^{+}(3^3P_1)$ &  6581 &  6650&  6557& 6792 & 6761 & 6516\\
        \hline
        $2^{+}(3^3P_2)$ &  6610 & 6678 &  6570& 6821 & 6780  & 6527\\
    \hline
   \end{tabular}
   \label{massresultsforbottommesons}
   \end{table}
\subsection{Decay Widths and Upper Bounds of Associated Couplings}
Using our calculated masses, we analyzed the widths of strong decays and predicted the upper bounds of the associated couplings. The strong decays of excited mesons involve the emission of $\pi, K,\eta$, which are treated as goldstone bosons; hence, it is convenient to analyze these interactions. We have studied decays of excited heavy-light mesons with emission of a pseudoscalar meson only. The decays with emissions of vector mesons ($\omega$, $\rho$, $K^{*}$, $\phi$) are also possible and studied in Ref.\cite{casalbuoni1992, schechter1993, campanella2018}. The contribution of decays with the emission of vector mesons into total decay widths is substantial for these mesonic states. The formulation for decay widths is discussed in Section 2. We apply the effective Lagrangian approach discussed in Section 2 to calculate OZI-allowed two body strong decay widths in terms of associated couplings.

Numerical values used for calculating decay width are $M_{\pi^{0}}$ = 134.97 $MeV$, $M_{\pi^{+}}$ = 139.57 $MeV$, $M_{K^{+}}$ = 493.67 $MeV$, $M_{\eta^{0}}$ = 547.85 $MeV$, $M_{K^{0}}$ = 497.61 $MeV$, $M_{B^{0}}$ = 5279.63 $MeV$, $M_{B^{*}}$ = 5324.65 $MeV$, $M_{B^{0}_{s}}$ = 5366.89 $MeV$, $M_{B^{*}_{s}}$ = 5415.40 $MeV$, and estimated masses for $n = 3$ bottom mesons states.

The computed strong decay widths in terms of $\dbtilde{g}_{HH}$, $\dbtilde{g}_{SH}$, $\dbtilde{g}_{TH}$ for radially excited $n = 3$ bottom mesons are presented in Table \ref{decaywidthsfornonstrangebottommesons}, \ref{decaywidthsforstrangebottommesons}, respectively. We also include suppression factors in decays, which arise due to the violation of isospin symmetry in decays. When the mass difference between the parent heavy-light meson with strangeness and the daughter non-strange heavy-light meson is less than the mass of kaons, then the breaking of isospin symmetry takes place, and the suppression factor $\epsilon$ is incorporated in the associated decay mode \cite{gross1979}. To account for isospin violation, the suppression factor is given by :
\begin{align}
    \epsilon^{2} = \frac{3}{16}\left(\frac{m_{d} - m_{u}}{m_{s}-(\frac{m_{u}+m_{d}}{2})}\right) \approx 10^{-4}
\end{align}
Here $m_{u}, m_{d}$, and $m_{s}$ are current quark masses. This suppression factor $\epsilon$ is multiplied with decay modes, which occur with isospin violation.

 \begin{longtable}[t]{c c c p{4cm} p{3.5cm} p{2cm}}
 \caption{ Decay widths of obtained masses for $n = 3$ bottom mesons.}\\    
 \toprule
 States & $J^{P}$ & Decay modes & Partial Decay widths ($MeV$) & Total decay width ($MeV$) \cite{godfrey2016b}& Upper bounds \\ \midrule \endhead \midrule
$B(6326)$ & $0^{-}$ & $B^{*}\pi^+$ & 6036.89$\dbtilde{g}_{HH}^2$&&\\
             &         & $B^{*}\pi^{0}$ & 3024.30$\dbtilde{g}_{HH}^2$ &&\\
             &         & $B^{*}\eta^{0}$& 2423.70$\dbtilde{g}_{HH}^2$&&\\
             &         & $B_{s}^{*}K^{0}$ & 2883.13 $\dbtilde{g}_{HH}^2$&&\\
             &      &Total & 17222.24$\dbtilde{g}_{HH}^2$ &150.6& \\
             &      &     & $\dbtilde{g}_{HH}$ && 0.09 \\
\hline
$B(6342)$ & $1^{-}$ & $B^{+}\pi^-$ &  2351.72$\dbtilde{g}_{HH}^2$&& \\
             &         &  $B^0\pi^0$ & 1178.60$\dbtilde{g}_{HH}^2$&& \\
             &         & $B^{*+}\pi^-$ & 2102.57$\dbtilde{g}_{HH}^2$&& \\
             &         & $B^{*0}\pi^0$ & 1053.26$\dbtilde{g}_{HH}^2$&& \\
             &         & $B^{0}\eta^0$ & 1010.34$\dbtilde{g}_{HH}^2$ &&\\
           
             &         & $B_{s}K^0$ & 1233.43$\dbtilde{g}_{HH}^2$ &&\\
          
             &         & $B_{s}^{*0}K^0$ & 1020.49$\dbtilde{g}_{HH}^2$&& \\
             &         &Total & 9950.41$\dbtilde{g}_{HH}^2$ &140.6&  \\
             &         & & $\dbtilde{g}_{HH}$ & &0.12 \\
\hline
$B(6568)$ &  $0^+$ & $B^{+}\pi^-$ &  11857.7$\dbtilde{g}_{SH}^2$&& \\
$B(6568)$             &        & $B^{0}\pi^0$ & 5925.86$\dbtilde{g}_{SH}^2$ &&\\
             &        & $B^{0}\eta^0$ & 6280.66$\dbtilde{g}_{SH}^2$&& \\
             &        & $B^{0}_{s}K^{0}$ &9420.99$\dbtilde{g}_{SH}^2$&& \\
             &        &Total & 33485.21$\dbtilde{g}_{SH}^2$&166.8&\\
             &        &      & $\dbtilde{g}_{SH}$ & &0.07 \\
\hline
     $B(6624)$ & $1^{+}$ & $B^{*0}\pi^0$ &  6084.5$\dbtilde{g}_{SH}^2$&& \\
       &  & $B^{*+}\pi^-$ &  12166.2$\dbtilde{g}_{SH}^2$ &&\\
      & & $B^{*0}\eta^0$ & 8445.9$\dbtilde{g}_{SH}^2$ &&\\
       & & $B^{*0}_{s}K^0$ & 9620.58$\dbtilde{g}_{SH}^2$&& \\
       & &Total & 36317.18$\dbtilde{g}_{SH}^2$&93.4&\\
       & & & $\dbtilde{g}_{SH}$ && 0.05 \\
       \hline
       $B(6581)$ & $1^{+}$ & $B^{*}\pi^+$ &  18708.7$\dbtilde{g}_{TH}^2$&& \\
       &  & $B^{*}\pi^0$ &  9373.41$\dbtilde{g}_{TH}^2$ &&\\
      & & $B^{*0}\eta^0$ & 9218.74$\dbtilde{g}_{TH}^2$ &&\\
       & & $B^{*0}_{s}K^0$ & 8432.98$\dbtilde{g}_{TH}^2$&& \\
       & &Total & 45733.83$\dbtilde{g}_{TH}^2$& 174.8 &\\
       & & & $\dbtilde{g}_{TH}$ && 0.06 \\
       \hline
       $B(6610)$ & $2^{+}$ & $B^{+}\pi^-$ &  9587.29$\dbtilde{g}_{TH}^2$&& \\
      & &  $B^0\pi^0$ & 4793.99$\dbtilde{g}_{TH}^2$ &&\\
      & & $B^{*+}\pi^-$ & 12394.20$\dbtilde{g}_{TH}^2$&& \\
       & & $B^{*0}\pi^0$ & 6209.21$\dbtilde{g}_{TH}^2$&& \\
       & & $B^{*0}\eta^0$ & 5130.64$\dbtilde{g}_{TH}^2$ &&\\
         & & $B_{s}^{0}K^0$ & 4756.24$\dbtilde{g}_{TH}^2$&& \\
        & & $B_{s}^{*0}K^0$ & 3850.11 $\dbtilde{g}_{TH}^2$ & &\\
          & & Total & 46721.68$\dbtilde{g}_{TH}^2$& 87.7&\\
          & & & $\dbtilde{g}_{TH}$ & & 0.043 \\ \bottomrule
          \label{decaywidthsfornonstrangebottommesons}
 \end{longtable}
\begin{longtable} [t]{c c c p{4cm} p{3.5cm} p{2cm}}
\caption{ Decay widths of obtained masses for $n = 3$ strange bottom mesons.}\\  
\toprule
States & $J^{P}$ & Decay modes & Partial Decay widths ($MeV$) & Total decay width ($MeV$) \cite{godfrey2016b} & Upper bounds \\ \midrule \endhead \midrule
$B_{s}(6463)$ & $0^{-}$ & $B^{*+}K^-$ & 6334.56$\dbtilde{g}_{HH}^2$&&\\
             &         & $B^{*0}K^{0}$ & 6299.04$\dbtilde{g}_{HH}^2$ &&\\
             &         & $B_{s}^{*}\eta^{0}$&4349$\dbtilde{g}_{HH}^2$&&\\
             &         & $B_{s}^{*}\pi^{0}$& 3434.59$\dbtilde{g}_{HH}^2$ $\times 10^{-4}$&&\\
             &      &Total & 33965.54$\dbtilde{g}_{HH}^2$ &120.80& \\
             &      &     & $\dbtilde{g}_{HH}$ && 0.06 \\
\midrule
$B_{s}(6474)$ & $1^{-}$ & $B^{+}K^-$ &  2461.5$\dbtilde{g}_{HH}^2$&& \\
             &         &  $B^{0}K^0$ & 2449.5$\dbtilde{g}_{HH}^2$&& \\
             &         & $B^{*+}K^-$ & 4363.97$\dbtilde{g}_{HH}^2$&& \\
             &         & $B^{*0}K^0$ & 4340.19$\dbtilde{g}_{HH}^2$&& \\
             &         & $B_{s}^{0}\eta^0$ & 295.15$\dbtilde{g}_{HH}^2$ &&\\
             &         & $B_{s}^{*}\pi^{0}$& 2357.22 $\dbtilde{g}_{HH}^2$ $\times 10^{-4}$&&\\
             &         & $B_{s}^{*}\eta^0$ & 503.26$\dbtilde{g}_{HH}^2$ &&\\
             &         &Total & 38777.78$\dbtilde{g}_{HH}^2$ &129.8&  \\
             &         & & $\dbtilde{g}_{HH}$ & &0.06 \\
\midrule
$B_{s}(6789)$ &  $0^+$ & $B^{+}K^-$ &  17188.00$\dbtilde{g}_{SH}^2$&& \\
             &        & $B^{0}K^0$ & 17168.00$\dbtilde{g}_{SH}^2$ &&\\
             &        & $B_{s}^{0}\eta^0$ & 2447.79$\dbtilde{g}_{SH}^2$&& \\
             &        & $B^{0}_{s}\pi^{0}$ &7676.52$\dbtilde{g}_{SH}^2$ $\times 10^{-4}$&& \\
             &        &Total & 70286.76$\dbtilde{g}_{SH}^2$&51&\\
             &        &      & $\dbtilde{g}_{SH}$ & &0.03 \\
\midrule
     $B_{s}(6826)$ & $1^{+}$ & $B_{s}^{*0}\pi^0$ &  7544.37$\dbtilde{g}_{SH}^2$ $\times 10^{-4}$&& \\
      & & $B_{s}^{*0}\eta^0$ & 2402.4$\dbtilde{g}_{SH}^2$ &&\\
       & & $B^{*0}K^0$ & 16993.60 $\dbtilde{g}_{SH}^2$&& \\
        & & $B^{*+}K^-$ & 17003.21 $\dbtilde{g}_{SH}^2$&& \\
       & &Total & 36399.96$\dbtilde{g}_{SH}^2$&108.2&\\
       & & & $\dbtilde{g}_{SH}$ && 0.05 \\
       \midrule
       $B_{s}(6792)$ & $1^{+}$ & $B^{*+}K^-$ &  27966.7$\dbtilde{g}_{TH}^2$&& \\
       &  & $B^{*0}K^0$ &  27822.9$\dbtilde{g}_{TH}^2$ &&\\
      & & $B_{s}^{*0}\pi^0$ & 14178.2$\dbtilde{g}_{TH}^2$ $\times 10^{-4}$ &&\\
       & & $B_{s}^{*0}\eta^0$ & 3132.08$\dbtilde{g}_{TH}^2$&& \\
       & &Total & 58923.09$\dbtilde{g}_{TH}^2$& 47.7 &\\
       & & & $\dbtilde{g}_{TH}$ && 0.03 \\
       \midrule
       $B_{s}(6821)$ & $2^{+}$ & $B^{*+}K^-$ &  18439.00$\dbtilde{g}_{TH}^2$&& \\
      & &  $B^{*0}K^0$ & 18348.20$\dbtilde{g}_{TH}^2$ &&\\
      & & $B_{s}^{*0}\pi^0$ & 9303.23$\dbtilde{g}_{TH}^2$ $\times 10^{-4}$&& \\
       & & $B_{s}^{*0}\eta^0$ & 2091.96$\dbtilde{g}_{TH}^2$&& \\
       & & $B^{+}K^-$ & 14139.4$\dbtilde{g}_{TH}^2$ &&\\
         & & $B^{0}K^0$ & 14058.10$\dbtilde{g}_{TH}^2$&& \\
        & & $B_{s}^{0}\pi^0$ & 7158.8 $\dbtilde{g}_{TH}^2$ $\times 10^{-4}$ & &\\
         & & $B_{s}^{0}\eta^0$ & 1655.24 $\dbtilde{g}_{TH}^2$ & &\\
          & & Total & 68733.35$\dbtilde{g}_{TH}^2$& 106.8&\\
          & & & $\dbtilde{g}_{TH}$ & & 0.04 \\
          \bottomrule
          \label{decaywidthsforstrangebottommesons}
\end{longtable}

The computed strong decay widths in terms of coupling constants $\dbtilde{g}_{HH}$, $\dbtilde{g}_{SH}$, $\dbtilde{g}_{TH}$ for $n = 3$ bottom mesons states are presented in Table \ref{decaywidthsfornonstrangebottommesons}, \ref{decaywidthsforstrangebottommesons}, respectively. Without enough experimental data, it is not possible to determine values of coupling constants from heavy quark symmetry solely, but the upper bounds to these coupling are mentioned in Table \ref{decaywidthsfornonstrangebottommesons}, \ref{decaywidthsforstrangebottommesons}. In our study, we are taking limited modes of decay and that also only to ground state. We believe that a particular state like $B(6326)$ give 17222.24 $\dbtilde{g}_{HH}^2$ total decay width; when compared with total decay widths mentioned by other theoretical paper \cite{godfrey2016,godfrey2016b}, we provided an upper bound on $\dbtilde{g}_{HH}$ value. Now if we take additional modes, then value of $\dbtilde{g}_{HH}$ will be lesser than 0.09 ($\dbtilde{g}_{HH}< 0.09$). So, these upper bounds may give important information to other associated bottom states. Large fractions of the decay width of any excited state are dominated by modes that include the ground state. This work also provides a lower limit to total decay width, giving important clues to forthcoming experimental studies. The weak and radiative decays are not included in the computed decay widths of charm and bottom mesons. We also exclude decays via emissions of vector mesons ($\omega,\rho, K^*,\phi$). They give the contribution of $\pm{50}$ $MeV$ \cite{godfrey2016,godfrey2016b} to total decay widths for states analyzed above.

The coupling constant plays an important role in hadron spectroscopy. They are coupled with decays and give information about the strength of strong transitions of excited heavy meson doublets into the highest heavy meson doublets. Here, dimensionless coupling constants $g_{HH}$, $g_{SH}$, $g_{TH}$, $g_{ZH}$, $g_{RH}$ give the strength of transitions between $H-H, S-H, T-H, Z-H$, and $R-H$ fields, respectively. Values of coupling constants are more for ground state transitions ($H-H$ fields) than excited states ($S-H, T-H, X-H, Z-H, R-H$ fields) transition shown by value of $g_{HH} =0.64\pm0.075$ \cite{colangelo2012} while $g_{SH} =0.56\pm 0.04$, $g_{TH} =0.43\pm 0.01$ \cite{colangelo2012}, $g_{XH} = 0.24$ \cite{wang2014}. Also, values of coupling constants are low at higher orders ($n = 2$, $n = 3$) in comparison to lower order ($n = 1$) interactions \cite{colangelo2012,casalbuoni1997} like $\tilde{g}_{HH} = 0.28\pm 0.015$, $\tilde{g}_{SH} = 0.18\pm 0.01$ so on. This progression also supports the values of coupling constants computed in this study.

\subsection{Regge Trajectories} 
The Regge trajectory approach is an extensively employed and effective technique for investigating hadron spectra. Plots of total angular momentum ($J$) and radial quantum number ($n_{r}$) of hadrons as a function of squared mass $M^{2}$ provide valuable insights into quantum no.s of specific states and serve as an effective tool for identifying newly observed hadronic states. The following definitions are employed throughout this work:

\begin{itemize}
\item[(a).]{The ($J$,$M^2$ ) Regge trajectories:
\begin{equation}
 J = \alpha M^2 + \alpha_{0}
\end{equation}}
\item[(b).]{The ($n_r$, $M^2$ ) Regge trajectories:
\begin{equation}
 n_r  = \beta M^2 + \beta_{0}
\end{equation} }
\end{itemize}
Here $\alpha$, $\beta$ are slopes and $\alpha_{0}$, $\beta_{0}$ are intercepts. We plot Regge trajectories in plane ($J$, $M^2$ ) with natural parity $P= (-1)^J$ and unnatural parity $P= (-1)^{J-1}$ for  heavy-light mesons using predicted spectroscopic data. The plots of Regge trajectories in the ($J, M^2$) plane are also known as Chew-Frautschi plots.

\begin{figure}[htp]
    \centering
    \includegraphics[width=5cm]{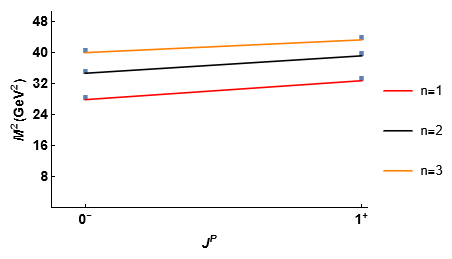}
    \caption{Regge trajectories with unnatural parity for non-strange bottom mesons.  }
    \label{Fig3.8}
\end{figure}
\begin{figure}[htp]
    \centering
    \includegraphics[width=5cm]{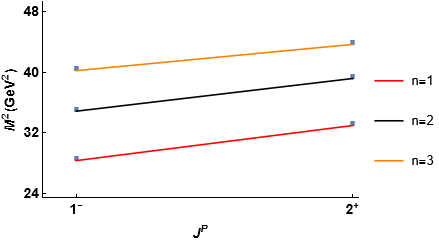}
    \caption{Regge trajectories with natural parity for non-strange bottom mesons.  }
    \label{Fig3.9}
\end{figure}
\begin{figure}[htp]
    \centering
    \includegraphics[width=5cm]{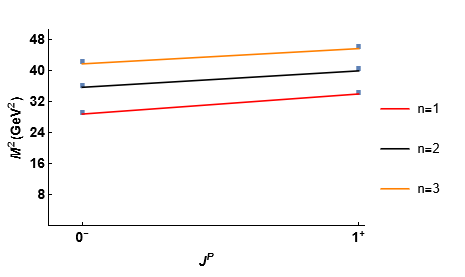}
    \caption{Regge trajectories with unnatural parity for strange bottom mesons.  }
    \label{Fig4}
\end{figure}
\begin{figure}[thp]
    \centering
    \includegraphics[width=5cm]{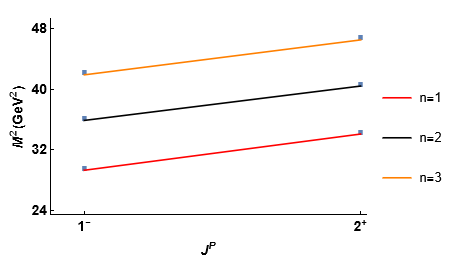}
    \caption{Regge trajectories with natural parity for strange bottom meson.  }
    \label{Fig4.1}
\end{figure}

\begin{figure}[htp]
    \centering
    \includegraphics[width=5cm]{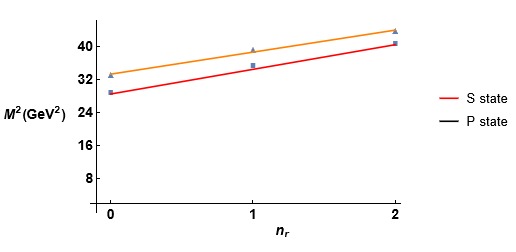}
    \caption{Regge trajectories for spin average masses for non strange bottom meson in plane($M^2\rightarrow n_r$)}
    \label{Fig4.2}
\end{figure}
\begin{figure}[htp]
    \centering
    \includegraphics[width=5cm]{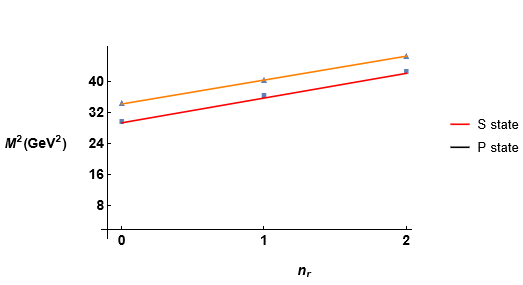}
    \caption{Regge trajectories for spin average masses for strange bottom meson in plane($M^2\rightarrow n_r$)}
    \label{Fig4.3}
\end{figure}
Regge trajectories in the plane ($n_r$, $M^2$) are constructed in Fig. $1 - 2$ using spin-averaged masses for $S$ and $P$-waves bottom mesons. In Fig. $1-2$, masses for $ n = 1$ are taken from PDG \cite{PDG2022}; for $ n = 2$, masses are taken from Ref. \cite{ebert2010}, and for $n = 3$, we are taking our calculated masses (Table 3). Using calculated slopes and intercepts, we predicted masses for $n = 4$ bottom mesons for both non-strange and strange states listed in Tables 6 and 7. 

\begin{table}[htp]
\centering
 \caption{Non strange bottom masses lying in Regge lines in plane $(n_{r}, M^{2})$.}
    \begin{tabular}{|c|c|c|c|}
    \hline
     State  & $J^{P}$ & Masses ($MeV$) & Ref.\cite{godfrey2016b} \\
     \hline
        $4^{1}S_{0}$ & $0^{-}$ & 6812& 6689\\
        \hline
        $4^{3}S_{1}$ & $1^{-}$ & 6811& 6703\\
        \hline
        $4^{3}P_{0}$ & $0^{+}$ & 6959& 6890\\
        \hline
         $4^{1}P_{1}$ & $1^{+}$& 7037& 6872\\
         \hline
        $4^{3}P_{1}$ & $1^{+}$& 6985& 6897\\ 
        \hline
        $4^{3}P_{2}$ & $1^{+}$& 7029& 6883\\
        \hline
    \end{tabular}
   
    \label{tab:my_label30}
\end{table}

\begin{table}[htp]
\centering
\caption{ Strange bottom masses Lying in Regge lines in $(n_{r},M^{2})$.}
    \begin{tabular}{|c|c|c|c|}
    \hline
     State  & $J^{P}$ & Masses ($MeV$) &Ref.\cite{godfrey2016b} \\
     \hline
        $4^{1}S_{0}$ & $0^{-}$ & 6958& 6759\\
        \hline
        $4^{3}S_{1}$ & $1^{-}$ & 6951& 6773\\
        \hline
        $4^{3}P_{0}$ & $0^{+}$ & 7217& 6950\\
        \hline
         $4^{1}P_{1}$ & $1^{+}$& 7254& 6946\\
         \hline
        $4^{3}P_{1}$ & $1^{+}$& 7224& 6959\\ 
        \hline
        $4^{3}P_{2}$ & $1^{+}$& 7265& 6956\\
        \hline
    \end{tabular}
    
    \label{tab:my_label31}
\end{table}

\section{Conculsion}
 Heavy quark symmetry serves as a fundamental framework for describing the spectroscopy of hadrons containing a single heavy quark. Utilizing available experimental and theoretical data on charm mesons, we employ heavy quark symmetry to predict the masses of $n = 3$ bottom mesons. Based on these computed masses, we investigate the decay widths of excited states transitioning to the ground state via the emission of pseudoscalar mesons. The decay widths are expressed in terms of coupling constants, which are determined by comparing our results with available theoretical predictions for total decay widths. These total decay widths provide an upper bound on the coupling constants, giving valuable insights into the properties of other associated bottom meson states. Regge trajectories are also constructed for our predicted data in planes ($J$, $M^2$ ) and ($n_r$, $M^2$ ), and estimate higher masses ($n = 4$) by fixing Regge slopes and intercepts. These findings may assist in the analysis of upcoming experimental results.
\section{Acknowledgement}
The authors gratefully acknowledge the financial support of the
Department of Science and Technology (SERB/F/9119/2020), New
Delhi and for Senior Research Fellowship (09/0677(11306)/2021-EMR-I) by Council of Scientific and Industrial Research, New Delhi. 

\bibliography{ref}
\bibliographystyle{epj}
\end{document}